# Exploit Kits: The production line of the Cybercrime Economy?

Michael Hopkins, Ali Dehghantanha

School of Computing, Science, and Engineering, University of Salford
Manchester, United Kingdom
michael.hopkins@project-clarity.com, A.Dehghantanha@Salford.ac.uk

*Abstract*— **The annual cost of Cybercrime to the global economy is estimated to be around $400 billion, in support of which Exploit Kits have been providing enabling technology since 2006. This paper reviews the recent developments in Exploit Kit capability and how these are being applied in practice. In doing so it paves the way for better understanding of the exploit kits economy that may better help in combatting them and considers industry's preparedness to respond.**

I. INTRODUCTION

Intel have estimated the annual cost of Cybercrime to the global economy to be around $400 billion[1]. Combating cyber crime remains a major research topic in multiple fields of digital forensics[2],[3], including applications forensics [4]–[8] mobile device forensics [9]–[11], cloud forensics [12]–[18], and malware investigation [19]–[25] . Despite this, the cybercrime rate continues to increase [26]! A growing contributor to this increase, since the introduction of one of the first examples "Mpack" around 2006[27], has been the Exploit Kit.

Defined by McAfee as "an off the shelf software package containing easy to use packaged attacks on known and unknown vulnerabilities"[28], the Exploit Kit serves an almost infinitely flexible menu of malware ranging from ransomware through banking and backdoor Trojans to rootkits.  By exploiting vulnerabilities, typically in web browsers and their plugins (Adobe Reader, Flash Player and Java etc.)[30]  the aim is often to achieve undetected remote control of the target device [31]. The arrest of the market leading providers in 2013[32], media reports of declining use and conflicting threat landscape surveys[33] are likely to lead to confusion as to their relative prevalence as a threat.

This paper aims to first provide an overview and contextualize recent developments in Exploit Kit capability, illustrates how these are being applied in practice, determine whether the overall trend is one of market growth or decline and finally discuss industry's preparedness to respond to exploit kits risks. The rest of this paper is organized as a Literature and Development Review followed by Discussion and some suggestions for Future Work.

II. LITERATURE AND DEVELOPMENT REVIEW

In December of 2013 ENISA, the Cyber Security body of the European Union, published a report on the evolving threat landscape[36]. Occupying positions number 1 and number 4 in the list respectively were "Drive by Exploits (malicious code injects to exploit web browser vulnerabilities)" and "Exploit Kits (ready to use software packages to automate cybercrime) ". Only two months prior, in October 2013 the alleged developer of the market leading Blackhole exploit kit, "Paunch", was arrested in a move by Russian authorities [37].  At this time, the Blackhole exploit kit was reputed to have around 50% market share, serving over one thousand customers.  The arrest was reported as a major blow to the cybercrime economy [32].

Just over 12 months later, at the end of January 2015, the impact of the Russian authorities action against the Blackhole providers seemed to be reflected in the declining threat of the exploit kit.  ENISA reported that Exploit kits, with an observed trend of "decreasing", occupied position 8 on the threat landscape [33]. Further consideration of observed activity during 2014 and 2015 does not however support the same "decreasing" conclusion.  Figure 1, provides a summary of exploit kit prevalance in 2013, 2014 and 2015 ([1],[28],[38]).

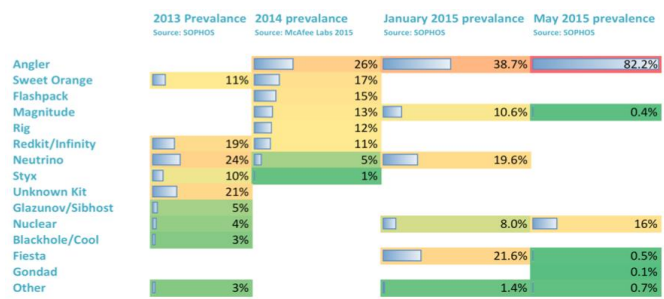

**Figure 1: Major Exploit Kit Providers 2013-2015**

As can be seen from Figure1 out of the exploit kits observed in 2013, only a minority persisted into 2014, with four new entrants (Angler, Flashpack, Magnitude, & Rig) accounting for around two thirds of market share in 2014. By 2015 the Paunch shaped hole in the exploit kit supply chain had been completely replaced by another dominant provider; the Angler exploit kit, which accounted for over 82% of observed attacks and infections.

Consideration of the exploit kit with reference to the Lockheed Martin Kill Chain[34][35], helps to understand the continuing appeal of the exploit kit and to provide some context to some of the most recent advances in exploit kit technology. By automating the majority of the first 5 steps from Reconnaisance to Installation with industrial scalability, an exploit kit enables the cyber criminal to instead focus attention on areas such as affiliate networking, monetization the and ongoing evasion of the command and control network.

Exploit kits are usually licensed using the software/"crimeware" as a service model, with clients being granted access to pre-imaged servers on which to load their binaries. In 2013 the Blackhole kit could be rented from $500 per month, its more sophisticated sister product "Cool" from $10,000 per month. In 2015 the RIG exploit kit is available from around $300 per month or by the day for less than $50 [39]. Grier et al.[27] further identified that instead of having to lure the victim and install their chosen malware themselves, the Traffic-PPI model, powered by the exploit kit, enabled clients to simply supply their binaries and pay per install. This business model is further supported by affiliates, whose referrals to the exploit landing sites are tracked and remunerated on a model within which the exploit kit provider typically pays per thousand visits.

Mirroring the legitimate software as a service economy, the deployment model gives exploit kit developers a high degree of control over their source code, how it is developed and supported. This approach will have facilitated much of the recent evolution whilst preserving the almost ethereal capability of today's kits to move between hosts, for the most part undetected.

It is therefore within Steps 2 and 5 (Weaponization & Installation) that the most significant changes have been observed in the last two years. A small selection of the developments include;

- **file-less intrusions**, where Windows directory traversal vulnerabilities have been exploited, co-opting Windows utilities to act for the exploit kit [40]

- **XOR encrypted payloads** being downloaded to the target machine which are then decrypted prior to execution[41]

- **Just in time assembly** of malware where a disassembled payload is secreted through perimeter defences and re-assembled on the client using Windows powershell[40].

- **a reduction in the average age of the vulnerabilities being targeted**, with many zero day exploits being woven into kits in 2015 [42],[43]. NSS labs reported that the average time for a zero day vulnerability to become known to software vendors and users is around 151 days [44] – providing a significant opportunity for integration and widespread exploitation within a kit. For these vulnerabilities, which can be sold for up to $1m, the exploit kit is likely to be the fastest route achieving a return on investment.

- **DNS hijacking** attacks directing users to exploit kits[45]

- **the use of TOR for communication** between infected hosts and the command and control servers[46], [47]

Although the use of dummy functions, manipulating strings, and obfuscating code with commercially available software are now widely used, kits such as Angler, Nuclear and Rig have further developed their Installation capability.

To avoid deployment on unsuitable platforms or sandboxed environments which may attract the unwanted attention of threat researchers, they seek first to identify the presence of virtual machines or anti-malware products which could frustrate payload delivery and/or expose operating practices.

Angler will look for the presence of security products (e.g. Kapersky, Trend Micro) and virtual machines (e.g. VMWare, Parallels). Within the Nuclear kit, this verification happens even before the target is redirected to the exploit kit landing page [46].

Developments have not been constrained only to Weaponisation and Installation, but have also been observed within Step 3 (Delivery). Although Phishing remains a successful vector for the Exploit kit, Malvertising delivers both targets and ongoing income for the cyber criminal in a perversely ingenious business model:

- Malicious adverts deployed by the cyber criminal lure potential targets to the exploit site, sometimes these "malverts" are procured and placed via third party networks, alternatively legitimate adverts are hacked
- Targets compromised by exploit kits are recruited as unsuspecting zombies to botnets
- These zombies generate fake page impressions on advertisers sites, delivering a click fraud revenue stream for the cyber criminal
- Click fraud proceeds fund Exploit Kit development and the next campaign

The US Association of National Advertisers projected that advertisers would lose $6.3 billion to bot related fraud in 2015[31]. Analysing 5.5 billion advertising impressions across 3 million domains over 60 days, almost one quarter of video advertising impressions were attributed to bot fraud. When traffic was sourced via a third party, bot fraud rates rose to over 50%. Prevention of these attacks presents serious challenges:

- The Malvert behaves in many ways identically to a legitimate advert – offering up embedded content and performing frequent, often nested and seamless redirects to content served by third party advertising networks
- As traffic is channelled through legitimate websites, web proxies blacklisting known bad sites provide little protection [40].
- The content delivered by third party networks is dynamic and constantly changing, so even forensic examination post infection is unlikely to reveal the same content as was delivered when the infection occurred

In the beginning of 2015, a threat actor, dubbed Fessleak by security firm Invincea [48], used a Malvertising vector alongside the Hanjuan exploit kit. The attack involved the Huffington post and an article relating to the Charlie Hebdo terrorist attack. Payloads included remote desktop capability and an ad-fraud bot. Fessleak was using both file-less flash and Adobe 0 day exploits to deliver his malware.

Finally considering Actions on Objectives, whilst the Exploit kit is becoming increasingly more sophisticated, McAfee report [1] the cyber criminal now needs fewer skills to take part in the industry. The Angler exploit kit does not require technical proficiency to launch an attack[28] and the cyber criminal is supported by initiatives including form based attack creation programs and affiliate schemes. Sophos further suggest the dominance by Angler may be related to a better return on investment for those criminals funding the kit developers (more traffic, better infection rates etc)[38].

In April 2015, it was observed that over 50% of the malware families installed by Angler were ransomware, including Teslacrypt, Kovter and Torrentlocker [38]. Attacks are not constrained to ransomware alone, in June 2015 a further malvertising attack was observed [49], this time using interstitial advertising (web pages displayed before or after an expected content page). The malvertising led to the Hanjuan exploit kit which was exploiting vulnerabilities in IE and flash to deliver a banking Trojan payload – the aim being to steal passwords and credentials via a man in the middle attack. The payload was itself encrypted and the kit used encryption between the command and control server.

In early October 2015, following a successful intervention, CISCO Talos attributed around 50% of the Angler Exploit Kit activity to one primary threat actor, reportedly generating more than $30m per annum by targeting up to 90,000 victims per day[50].

III. DISCUSSION AND FUTURE WORK

It is clear that far from being in decline, the Exploit kits have been on an accelerated evolutionary path. The last two years have seen the replacement of most Exploit Kits on the market and the emergence of an even more dominant player than was apparent in 2013 with the rise of the Angler kit [50].

The kits are more sophisticated than their predecessors with significant innovations in Weaponisation, Exploitation and Installation. Their automated delivery capability is ideally suited to mass market vectors such as Malvertising and has clearly powered the growth in click fraud and ransomware.

Those at the heart of this business have achieved Actions on Objectives on a scale which can only be described as industrial, generating revenues which eclipse most legitimate businesses[50]. It is also apparent that, surrounding the exploit kit provider, is an opaque yet complex ecosystem of actors which includes exploit providers, malvertising agents, affiliates and crime gangs. Whilst there are some excellent academic works, most current intelligence is emerging from the security industry, as attacks are discovered and analysed. This same industry however appears to be offering numerous, disparate solutions and is far from consensus on the most appropriate strategies for prevention and cure.

Advice includes selecting internet service providers with strong phishing and spam defenses, not opening unexpected attachments and using browser plugins to prevent the execution of scripts and iframes. As some security insiders have elected to completely remove programs such as Flash [51], other vendors believe the perimeter focused, signature based detection approach is already flawed and we must instead turn our capability towards isolation, containment and observation[40].

Reiterating the continued requirement for proactivity towards areas such as configuration management, patching and password security in order to reduce the attack surface, they also report that both business and consumers remain surprisingly inattentive[1]. In the midst of this apparent confusion and apathy, ENISA further identified exploit kits as emerging threat number 8 in the mobile computing environment and position 9 in the list of emerging threats to the "Trust Infrastructure" [33]. The latter is the code and data we use to ensure trusted connections when we communicate (encryption, digital signatures, challenge/response, SSL…).

An accurate appreciation of the threat posed by the exploit kit and the ways in which this could be addressed will require further clarity on their prevalance, their capability and surrounding ecosystem(s). What is certainly clear is that the exploit kit is bringing increasing sophistication to a cybercrime battle being waged on many fronts, together with the potential to extend the range and reach of the cyber criminal on an industrial scale.

## References

[1] "McAfee Labs Threat Report - August 2015." [Online]. Available: http://www.mcafee.com/us/resources/reports/rp-quarterly-threats-aug-2015.pdf. [Accessed: 19-Oct-2015].
[2] F. Daryabar, A. Dehghantanha, N. I. Udzir, N. F. Mohd Sani, S. Shamsuddin, and F. Norouzizadeh, "A survey on privacy impacts of digital investigation," J. Gener. Inf. Technol., vol. 4, no. 8, pp. 57–68, 2013.
[3] A. Dehghantanha and K. Franke, "Privacy-respecting digital investigation," in Privacy, Security and Trust (PST), 2014 Twelfth Annual International Conference on, 2014, pp. 129–138.
[4] M. N. Yusoff, R. Mahmod, M. T. Abdullah, and A. Dehghantanha, "Performance Measurement for Mobile Forensic Data Acquisition in Firefox OS," Int. J. Cyber-Secur. Digit. Forensics IJCSDF, vol. 3, no. 3, pp. 130–140, 2014.
[5] M. Ibrahim and A. Dehghantanha, "Modelling based approach for reconstructing evidence of VoIP malicious attacks," Int. J. Cyber-Secur. Digit. Forensics IJCSDF, vol. 3, no. 4, pp. 183–199, 2014.
[6] M. N. Yusoff, R. Mahmod, A. Dehghantanha, and M. T. Abdullah, "Advances of mobile forensic procedures in Firefox OS," Int. J. Cyber-Secur. Digit. Forensics IJCSDF, vol. 3, no. 4, pp. 183–199, 2014.
[7] J. Talebi, A. Dehghantanha, and R. Mahmoud, "Introducing and analysis of the Windows 8 event log for forensic purposes," in Computational Forensics, Springer International Publishing, 2015, pp. 145–162.
[8] F. Norouzizadeh Dezfouli, A. Dehghantanha, B. Eterovic-Soric, and K.-K. R. Choo, "Investigating Social Networking applications on smartphones detecting Facebook, Twitter, LinkedIn and Google+ artefacts on Android and iOS platforms," Aust. J. Forensic Sci., pp. 1–20, 2015.
[9] S. Mohtasebi, A. Dehghantanha, and H. G. Broujerdi, "Smartphone Forensics: A Case Study with Nokia E5-00 Mobile Phone," Int. J. Digit. Inf. Wirel. Commun. IJDIWC, vol. 1, no. 3, pp. 651–655, 2011.
[10] S. H. Mohtasebi and A. Dehghantanha, "Towards a Unified Forensic Investigation Framework of Smartphones," Int. J. Comput. Theory Eng., vol. 5, no. 2, p. 351, 2013.
[11] M. N. Yusoff, R. Mahmod, M. T. Abdullah, and A. Dehghantanha, "Mobile forensic data acquisition in Firefox OS," in Cyber Security, Cyber Warfare and Digital Forensic (CyberSec), 2014 Third International Conference on, 2014, pp. 27–31.
[12] M. Damshenas, A. Dehghantanha, R. Mahmoud, and S. bin Shamsuddin, "Cloud Computing and Conflicts with Digital Forensic Investigation," Int. J. Digit. Content Technol. Its Appl., vol. 7, no. 9, p. 543, 2013.
[13] M. Shariati, A. Dehghantanha, B. Martini, and K. R. Choo, "Ubuntu One investigation: Detecting evidences on client machines," 2015.
[14] M. Shariati, A. Dehghantanha, and K.-K. R. Choo, "SugarSync forensic analysis," Aust. J. Forensic Sci., no. ahead-of-print, pp. 1–23, 2015.
[15] F. Daryabar and A. Dehghantanha, "A review on impacts of cloud computing and digital forensics," Int. J. Cyber-Secur. Digit. Forensics IJCSDF, vol. 3, no. 4, pp. 183–199, 2014.
[16] F. Daryabar, A. Dehghantanha, N. I. Udzir, and others, "A Review on Impacts of Cloud Computing on Digital Forensics," Int. J. Cyber-Secur. Digit. Forensics IJCSDF, vol. 2, no. 2, pp. 77–94, 2013.
[17] M. Damshenas, A. Dehghantanha, R. Mahmoud, and S. Bin Shamsuddin, "Forensics investigation challenges in cloud computing environments," in Cyber Security, Cyber Warfare and Digital Forensic (CyberSec), 2012 International Conference on, 2012, pp. 190–194.
[18] A. Aminnezhad, A. Dehghantanha, M. T. Abdullah, and M. Damshenas, "Cloud Forensics Issues and Opportunities," Int J Inf Process Manag, vol. 4, no. 4, 2013.
[19] S. bin Shamsuddin and F. Norouzizadeh, "Analysis of Known and Unknown Malware Bypassing Techniques," 2013.
[20] M. Damshenas, A. Dehghantanha, K.-K. R. Choo, and R. Mahmud, "M0droid: An android behavioral-based malware detection model," J. Inf. Priv. Secur., vol. 11, no. 3, pp. 141–157, 2015.
[21] F. N. Dezfouli, A. Dehghantanha, R. Mahmod, N. F. B. M. Sani, S. B. Shamsuddin, and F. Daryabar, "A Survey on Malware Analysis and Detection Techniques," Int. J. Adv. Comput. Technol., vol. 5, no. 14, p. 42, 2013.
[22] M. Damshenas, A. Dehghantanha, and R. Mahmoud, "A survey on malware propagation, analysis, and detection," Int. J. Cyber-Secur. Digit. Forensics IJCSDF, vol. 2, no. 4, pp. 10–29, 2013.
[23] F. Daryabar, A. Dehghantanha, and H. G. Broujerdi, "Investigation of malware defence and detection techniques," Int. J. Digit. Inf. Wirel. Commun. IJDIWC, vol. 1, no. 3, pp. 645–650, 2011.
[24] S. Mohtasebi and A. Dehghantanha, "A Mitigation Approach to the Malwares Threats of Social Network Services," Muktimedia Inf. Netw. Secur., pp. 448–459, 2009.
[25] K. Shaerpour and A. Dehghantanha, "Trends in android malware detection," J. Digit. Forensics Secur. Law, vol. 8, no. 3, p. 21, 2013.
[26] M. Ganji, A. Dehghantanha, N. IzuraUdzir, and M. Damshenas, "Cyber Warfare Trends and Future," Adv. Inf. Sci. Serv. Sci., vol. 5, no. 13, p. 1, 2013.

[27] C. Grier, L. Ballard, J. Caballero, N. Chachra, C. J. Dietrich, K. Levchenko, P. Mavrommatis, D. McCoy, A. Nappa, A. Pitsillidis, N. Provos, M. Z. Rafique, M. A. Rajab, C. Rossow, K. Thomas, V. Paxson, S. Savage, and G. M. Voelker, "Manufacturing Compromise: The Emergence of Exploit-as-a-service," in Proceedings of the 2012 ACM Conference on Computer and Communications Security, New York, NY, USA, 2012, pp. 821–832.
[28] "McAfee Labs Threat Report - February 2015." [Online]. Available: http://www.mcafee.com/uk/resources/reports/rp-quarterly-threat-q4-2014.pdf. [Accessed: 18-Oct-2015].
[30] J. Chen and L. Brooks, "Evolution of Exploit Kits," Trend Micro. [Online]. Available: https://www.trendmicro.com/cloud-content/us/pdfs/security-intelligence/white-papers/wp-evolution-of-exploit-kits.pdf.
[31] "ANA/White Ops Study Reveals Extent of Advertising Bot Fraud | About the ANA | ANA." [Online]. Available: https://www.ana.net/content/show/id/32948. [Accessed: 18-Oct-2015].
[32] "Meet Paunch: The Accused Author of the BlackHole Exploit Kit — Krebs on Security." [Online]. Available: http://krebsonsecurity.com/2013/12/meet-paunch-the-accused-author-of-the-blackhole-exploit-kit/. [Accessed: 19-Oct-2015].
[33] "ENISA Threat Landscape 2014 — ENISA." [Online]. Available: https://www.enisa.europa.eu/activities/risk-management/evolving-threat-environment/enisa-threat-landscape/enisa-threat-landscape-2014. [Accessed: 15-Oct-2015].
[34] E. Hutchins, M. Cloppert, and R. Amin, "Intelligence-Driven Computer Network Defense Informed by Analysis of Adversary Campaigns and Intrusion Kill Chains." [Online]. Available: http://www.lockheedmartin.co.uk/content/dam/lockheed/data/corporate/documents/LM-White-Paper-Intel-Driven-Defense.pdf.
[35] "LM-White-Paper-Intel-Driven-Defense.pdf." .
[36] "ENISA Threat Landscape 2013 - Overview of current and emerging cyber-threats — ENISA." [Online]. Available: https://www.enisa.europa.eu/activities/risk-management/evolving-threat-environment/enisa-threat-landscape/enisa-threat-landscape-2013-overview-of-current-and-emerging-cyber-threats. [Accessed: 15-Oct-2015].
[37] "Blackhole exploit kit author arrested in Russia." [Online]. Available: http://www.pcworld.com/article/2053180/blackhole-exploit-kit-author-arrested-in-russia.html. [Accessed: 15-Oct-2015].
[38] "A closer look at the Angler exploit kit | Sophos Blog." [Online]. Available: https://blogs.sophos.com/2015/07/21/a-closer-look-at-the-angler-exploit-kit/. [Accessed: 22-Oct-2015].
[39] MalwareTech, "RIG Exploit Kit - Source Code Leak | MalwareTech." [Online]. Available: http://www.malwaretech.com/2015/02/rig-exploit-kit-possible-source-code.html. [Accessed: 22-Oct-2015].
[40] "The First Law of Intrusion Detection: You Can't Detect What You Can't See," Invincea. [Online]. Available: https://www.invincea.com/2015/09/the-first-law-of-intrusion-detection-you-cant-detect-what-you-cant-see/. [Accessed: 19-Oct-2015].
[41] "Malware Injected Directly Into Processes in Angler Exploit Kit Attack | SecurityWeek.Com." [Online]. Available: http://www.securityweek.com/malware-injected-directly-processes-angler-exploit-kit-attack. [Accessed: 02-Nov-2015].
[42] "HanJuan EK fires third Flash Player 0day," Malwarebytes Unpacked. [Online]. Available: https://blog.malwarebytes.org/exploits-2/2015/02/hanjuan-ek-fires-third-flash-player-0day/. [Accessed: 17-Oct-2015].
[43] "New Ransomware, FessLeak, Taps Adobe Flash Flaws," The Security Ledger. [Online]. Available: https://securityledger.com/2015/02/new-ransomware-fessleak-taps-adobe-flash-flaws/. [Accessed: 02-Nov-2015].
[44] "Hacking The Zero-Day Vulnerability Market," Dark Reading. [Online]. Available: http://www.darkreading.com/vulnerability/hacking-the-zero-day-vulnerability-marke/240164591. [Accessed: 19-Oct-2015].
[45] "Attackers Use Exploit Kit to Hijack Routers: Researcher | SecurityWeek.Com." [Online]. Available: http://www.securityweek.com/attackers-use-exploit-kit-hijack-routers-researcher. [Accessed: 02-Nov-2015].
[46] "Exploit Kits Improve Evasion Techniques," McAfee. [Online]. Available: https://blogs.mcafee.com/mcafee-labs/new-exploit-kits-improve-evasion-techniques/. [Accessed: 18-Oct-2015].
[47] "Shade among top three encryptors in Russia delivered via spam, exploit kits," SC Magazine. [Online]. Available: http://www.scmagazine.com/news/shade-encryptor-threat-in-russia-ukraine-germany/article/438463/. [Accessed: 02-Nov-2015].
[48] "Fessleak: The Zero-Day Driven Advanced RansomWare Malvertising Campaign," Invincea. [Online]. Available: https://www.invincea.com/2015/02/fessleak-the-zero-day-driven-advanced-ransomware-malvertising-campaign/. [Accessed: 17-Oct-2015].
[49] "Elusive HanJuan EK Drops New Tinba Version (updated)," Malwarebytes Unpacked. [Online]. Available: https://blog.malwarebytes.org/intelligence/2015/06/elusive-hanjuan-ek-caught-in-new-malvertising-campaign/. [Accessed: 17-Oct-2015].
[50] "Talos Intel - Threat Spotlight: Angler Exposed Generating Millions in Revenue." [Online]. Available: http://talosintel.com/angler-exposed/. [Accessed: 22-Oct-2015].
[51] "Angler Exploit Kit — Krebs on Security." [Online]. Available: http://krebsonsecurity.com/tag/angler-exploit-kit/. [Accessed: 19-Oct-2015].